\documentclass[prd,tightenlines,nofootinbib,superscriptaddress,twocolumn]{revtex4}
\usepackage[colorlinks=true,linkcolor=blue,citecolor=magenta,linktocpage=true]{hyperref}
\usepackage{mathtools}
\usepackage{amssymb}
\usepackage{ORCIDinREVTeX}

\begin{document}
\begin{flushright} {\footnotesize YITP-26-12, RESCEU-4/26, IPMU26-0003}  \end{flushright}
\title{The emergent Big Bang scenario}
\author{{\bf Justin C. Feng}}
\orcid{0000-0003-2441-5801}
\email{feng@fzu.cz}
\affiliation{
  Central European Institute for Cosmology and Fundamental Physics
  (CEICO), Institute of Physics of the Czech Academy of Sciences, Na
  Slovance 1999/2, 182 21 Prague 8, Czech Republic
}

\author{{\bf Shinji Mukohyama}}\orcid{0000-0002-9934-2785}
\email{shinji.mukohyama@yukawa.kyoto-u.ac.jp}
\affiliation{Center for Gravitational Physics and Quantum Information, Yukawa Institute for Theoretical Physics, Kyoto University, 606-8502 Kyoto, Japan}
\affiliation{Research Center for the Early Universe (RESCEU), Graduate School of Science, The University of Tokyo, Hongo 7-3-1, Bunkyo-ku, Tokyo 113-0033, Japan}
\affiliation{Kavli Institute for the Physics and Mathematics of the Universe (WPI), The University of Tokyo Institutes for Advanced Study (UTIAS), The University of Tokyo, Kashiwa, Chiba 277-8583, Japan}

\author{{\bf Jean-Philippe Uzan}}
\orcid{0009-0002-3927-5477}
\email{uzan@iap.fr}
\affiliation{Institut d’Astrophysique de Paris, UMR-7095 du CNRS, 98 bis boulevard
Arago, 75014 Paris, France}
\affiliation{Center for Gravitational Physics and Quantum Information, Yukawa Institute for Theoretical Physics, Kyoto University, 606-8502 Kyoto, Japan}
\begin{abstract}
This paper proposes a new avenue for understanding the cosmological singularity. The standard cosmological model contains a generic initial singularity usually referred to as the {\em big bang}. Herein, we present a novel idea to extend the description of our Universe beyond this limit. The proposal relies on rewriting physics in a purely Riemannian, {\em i.e.} locally Euclidean, four-dimensional space and the emergence of Lorentzian patches owing to the interaction of all matter fields to a clock field that is responsible for a signature change. If our Universe is contained within one of these patches, the initial singularity is replaced by a smooth boundary on which the signature of the physical metric flips. In this paper, we first define the model and draw the necessary conditions on its arbitrary functions for solutions to exist. Next, we prove the existence of solutions that lead to an emergent universe with a primordial (almost) de Sitter phase. To finish, we discuss the consequences of this construction for the universe on scales much larger than our observable Universe: a large ``Euclidean sea'' in which Lorentzian islands locally emerge and host an expanding universe potentially similar to ours. While speculative, this scenario has specific features that can be tested, and the present paper sets the basis for further phenomenological investigations.
\end{abstract}
\maketitle
\section{Introduction} \label{section1}

The existence of the primordial singularity of the standard cosmological model has long been identified as a signal of the domain of validity of general relativity to describe gravitation, starting perhaps with Lema\^{\i}tre's early insight~\cite{Lemaitre:1931zzb}. Many classical solutions to avoid the singularity within general relativity have been proposed, such as the Eddington-Lema\^{\i}tre model~\cite{Eddington:1930zz,Lemaitre:1931zza}, bouncing solutions that both require a nonvanishing spatial curvature, {\em i.e.} an effective fluid with equation of state $P=-\rho/3$ with positive or negative energy density according to the sign of the spatial curvature, oscillating universes~\cite{Tolman:1931fei,Lemaitre:1933gd,Ellis:2002we} or even a continuous creation of matter~\cite{Hoyle:1948zz}. Indeed those solutions either are not compatible with the actual measurements of the cosmological parameters or do not avoid the singularity.

The occurrence of singularities in general relativity has been related to energy conditions~\cite{Hawking:1966vg,Hawking:1966sx,Hawking:1973uf} and those early classical solutions have all seen counterparts in the developments of quantum gravity phenomenology, including bouncing universes thanks to duality~\cite{Veneziano:1991ek,Brandenberger:1988aj}, the dynamics of the dilaton~\cite{Gasperini:1992em}, branes~\cite{Steinhardt:2002ih,Steinhardt:2001st} in string theory, or the  the quantum pressure~\cite{Rovelli:1997yv,Bojowald:2005epg} that appears in loop quantum gravity. On the side of quantum cosmology, many solutions, invoking the no-boundary proposal~\cite{Hartle:1983ai}, tunneling geometries~\cite{Vilenkin:1982de}, or phase transitions in spin network models~\cite{Mielczarek:2012pf} inspired by signature change in metamaterials~\cite{Smolyaninov:2010tpq}, have been considered to describe the emergence of our Universe. 

It is worth noting that one can find solutions of the Einstein equations for a metric with Lorentzian as well as Euclidean signatures. It was actually conjectured by Sakharov~\cite{Sakharov:1984csx}, inspired by Ref.~\cite{Vilenkin:1982de}, that ``there may exist states of the physical continuum which include regions with different signatures of the metric and that the observed Universe and an infinite number of other universes arose as a result of quantum processes with a change in the signature of the metric.'' This led several authors~\cite{Ellis:1991sp,Ellis:1991st,Hayward:1992zp,Dray:1996cw,Dray:1996dc} to study the junction conditions between such spacetime patches with different signatures at the classical level, which require these regions to be glued smoothly on their boundaries with an either smooth~\cite{Ellis:1991sp} or discontinuous~\cite{Ellis:1991st,Hayward:1992zp} sign shift. However, these studies assume a profile of sign changing $g_{00}$ and do not construct a full model of time emergence based on a consistent field theory as proposed in Refs.~\cite{Mukohyama:2013ew,Kehayias:2014uta}---see also earlier work in the context of Nordstr\"om gravity \cite{Girelli:2008qp}.

Hence, following our previous investigations~\cite{Mukohyama:2013ew,Kehayias:2014uta,Mukoyama:2013gqu,Mukohyama:2013gra,Uzan-CUP,Feng:2023klt,Feng:2025xsi} on a field theory based on a purely Euclidean (hereafter, by ``Euclidean'' we mean ``Riemannian,'' {\em i.e.} ``locally Euclidean'') space, we propose in this paper a new road to understanding the big bang singularity. This theoretical approach relies on the hypothesis that ({\em i}) at the microscopic level the metric is Riemannian, {\em i.e.}, that all matter fields are defined in a four-dimensional Euclidean space with metric of signature $(+,+,+,+)$, 
\begin{eqnarray}
 \mathrm{d} s^2_{\rm E} &=& g^{\rm E}_{\mu\nu} \mathrm{d} x^\mu \mathrm{d} x^\nu,
 \end{eqnarray}
and ({\em ii}) interact with a {\em clock field} $\phi$ through its gradient so that they are at the macroscopic level universally coupled to the effective metric
\begin{equation}\label{e.shiftsignature}
g_{\mu\nu}= g^{\rm E}_{\mu\nu}-\frac{\partial_\mu\phi \partial_\nu\phi}{M^4}\, ,
\end{equation}
where $M$ is a mass scale. References~\cite{Mukohyama:2013ew,Kehayias:2014uta,Mukoyama:2013gqu,Mukohyama:2013gra,Uzan-CUP,Feng:2023klt,Feng:2025xsi}  demonstrated that if the gradient of the clock field develops a constant nonvanishing vacuum expectation value on a patch ${\cal M}_0$, then {\em within this patch} the effective metric can exhibit a $(-,+,+,+)$ signature. Furthermore, it was shown that the whole action of the standard model of particle physics in the patch ${\cal M}_0$ can be obtained while gravity is described by a more general covariant Galileon theory. While in the standard approach of  any field theory, the Lorentzian metric structure allows one to implement the relativistic notion of causality and to define a notion of time dimension, it is now seen as an emergent property. Hence, in this description, there is no dynamics nor signature flip across some hypersurface; instead, all the fields develop a Lorentzian dynamics in some patches of space because they minimally couple to an effective metric that depends on the local value of the clock field.

This insight offers a new view on the notion of time as an emergent property. From a phenomenological point of view, this shall affect the construction of our cosmological model since the whole dynamical observable Universe with a well-defined time arrow can exist only within the patch ${\cal M}_0$. On the boundaries $\Sigma_0$ of ${\cal M}_0$, ``time freezes'' since the space becomes Euclidean but without necessary passing through a singularity. Hence, one is led to the idea that if our Universe is contained in such a patch ${\cal M}_0$, then the cosmological singularity of the big bang is replaced by the boundary $\Sigma_0$  of this patch. 

\begin{figure}[htb]
 	\centering
 	\includegraphics[width=0.5\textwidth]{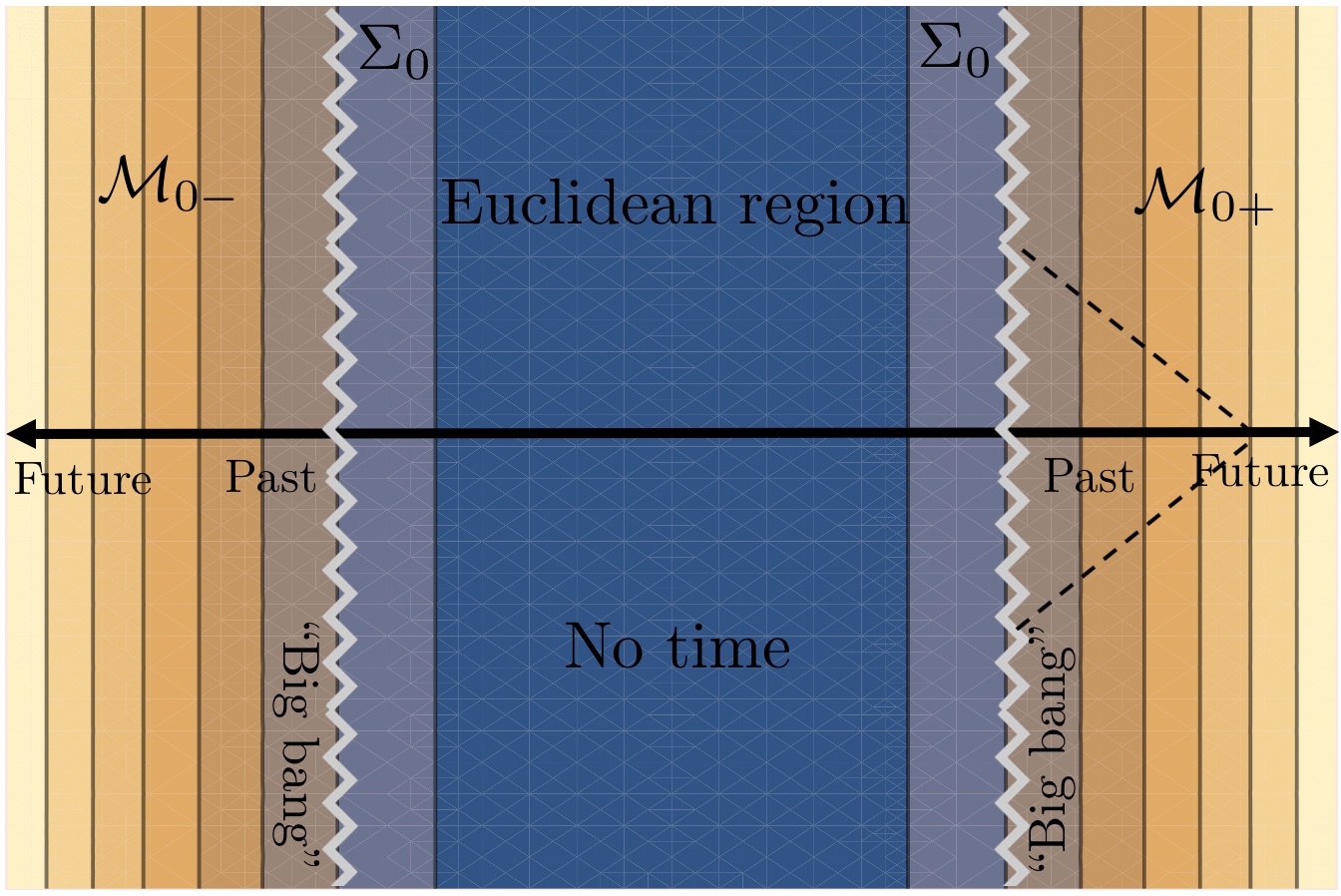} 
 	\caption{A big bang that never happened. Our Universe can be thought as an emergent Lorentzian patch within a four-dimensional purely Riemannian space. The ``big bang'' can be understood as the hypersurface $\Sigma_0$ on which the signature of the metric flips. In the simplest version, the geometry and field configurations are assumed to ({\em i}) depend only on one space coordinate, $z$, so that the emergent spacetimes ${\cal M}_{0\pm}$ satisfy the Copernican principle and ({\em ii}) exhibit the $(-,+,+,+)$ signature for the effective metric in the two asymptotic regions.}
 	\label{fig:1}
 \end{figure}

To investigate this idea further and make it more concrete, we shall construct a four-dimensional Euclidean space with planar symmetry such that both $g^{\rm E}_{\mu\nu}$ and $\phi$ depend only on one coordinate, $z$, so that the $\lbrace z={\rm const.}\rbrace$ hypersurfaces are homogeneous and isotropic, ensuring that the emergent universe\footnote{To avoid ambiguities, we shall use the term {\em universe} for the Lorentzian regions for which we can use the words ``time'' and ``space'' in a well-defined way.} satisfies the Copernican principle. Indeed, if a shift of signature~(\ref{e.shiftsignature}) occurs within ${\cal M}_0$ the $z$ direction will play the role of the arrow of time since $\lbrace z={\rm const.}\rbrace$  will be $\lbrace \phi={\rm const.}\rbrace$ and thus $\lbrace t={\rm const.}\rbrace$. We further assume that the effective metric has the signature $(-,+,+,+)$ in the asymptotic regions. The whole global space structure is as described in Fig.~\ref{fig:1}; that is, two ``mirror'' universes ${\cal M}_{0+}$ and ${\cal M}_{0-}$ with signature $(-,+,+,+)$ match a central Euclidean region with no time. Each of the two Lorentzian zones can be thought as a universe, with time flowing in opposite directions. What the observers in each emergent universe will identify as the big bang is simply the boundary $\Sigma_0$ on which the signature of the metric flips.

\section{Emergent big bang model}\label{section2}

\subsection{Definition of the theory}\label{subsec2.1}

We decompose the Euclidean metric $g^{\rm E}_{\mu\nu}$ in terms of the lapse $N_E$ and the metric $\gamma_{ij}$ on the three-dimensional hypersurface normal to a coordinate $z$ as
\begin{eqnarray} \label{e.met1}
 \mathrm{d} s^2_{\rm E} = N_{\rm E}^2(z) \mathrm{d} z^2 + \hbox{e}^{2h(z)}\delta_{ij}\mathrm{d} x^i\mathrm{d} x^j
\end{eqnarray}
with
\begin{equation}\label{e.spacegeo}
\gamma_{ij}(z) = \hbox{e}^{2h(z)}\delta_{ij}=a^2(z)\delta_{ij} \,.
\end{equation} 
Indeed, $h=\ln a$ so that $\dot h ={\dot a}/{a}$ with a dot denoting a derivative with respect to $z$, and for later convenience we define
\begin{equation}
 H_{\rm E} \equiv \dot{h}/N_{\rm E}\, .
\end{equation} 
The gradient of the clock field $\phi(z)$ is given by
\begin{equation}
 \psi(z)={\dot \phi }/{N_{\rm E}} \, ,
\end{equation}
so that the configuration up to a constant shift of $\phi$ is characterized by the three functions $N_{\rm E}$, $h$ and $\psi$ of the variable $z$.  $N_{\rm E}$ can always be set to unity thanks to a redefinition of $z$ so that we only have effectively two functions to determine the model.

Following Ref.~\cite{Mukohyama:2013ew}---see its  Eq.~(3.6)---we consider the action $S = \int {\cal L} \mathrm{d}^4x$ with Lagrangian\footnote{In \cite{Mukoyama:2013gqu} (cf. Sec. IV of \cite{Mukohyama:2013gra}), it was shown that on the Lorentzian patch ${\cal M}_0$, ${\cal L}$ may be rewritten in the Lorentzian form
${\cal L}_{\rm L}\! = \! \sqrt{-g} \left\lbrace f R\!  +\!  P\! + \!  2 f'\left[(\nabla^2\phi)^2\!  -\! (\nabla_\mu\nabla_\nu\phi)^2\right]\right\rbrace$, where  $R$ and $\nabla_\mu$ are the respective Ricci scalar and Levi-Civita connection for $g_{\mu\nu}$ given by \eqref{e.shiftsignature} and $f=f(X)$ and $P=P(X)$ are functions satisfying $f/G_4=P/{\cal K}=\sqrt{X/X_{\rm E}}$, where $X=-g^{\mu\nu}\partial_\mu\phi\partial_\mu\phi=(1/M^4-1/X_{\rm E})^{-1}$.
}
\begin{eqnarray}\label{e.S1}
{\cal L}\! &=&\! \sqrt{g_{\rm E}} \left\lbrace G_4 R_{\rm E}\!  +\!  {\cal K}\! -\!  2 G_4'\left[(\nabla^2_{\rm E}\phi)^2\!  -\! (\nabla_\mu^{\rm E}\nabla_\nu^{\rm E}\phi)^2\right]\right\rbrace 
\end{eqnarray}
with $\nabla^{\rm E}_\mu$ the covariant derivative associated to $g^{\mu\nu}_{\rm E}$, $\nabla^2_{\rm E}\phi \equiv g^{\mu\nu}_{\rm E}\nabla_\mu\nabla_\nu\phi = (\dot \psi+ 3 \dot h \psi) /N_{\rm E}$, and $\left(\nabla_\mu\nabla_\nu\phi\right)^2 \equiv g^{\mu\sigma}_{\rm E} g^{\nu\tau}_{\rm E} \nabla_\mu\nabla_\nu\phi \nabla_\sigma\nabla_\tau\phi$. The action involves only the two free functions, ${\cal K}$ and $G_4$, that depend only on $X_{\rm E}$ defined as
\begin{equation}
 X_{\rm E} \equiv  g^{\mu\nu}_{\rm E}\partial_\mu\phi \partial_\nu\phi =  \left(\frac{\dot\phi}{N_{\rm E}}\right)^2 \equiv \psi^2,
\end{equation} 
and we shall use a prime to denote a derivative with respect to $X_{\rm E}$, {\em i.e.} $G_4'\equiv \mathrm{d} G_4/\mathrm{d} X_{\rm E}$.  

Since the Lagrangian does not depend explicitly on $\phi$, the equation of motion for the scalar field takes the form of a current conservation equation; $\partial_\mu J^\mu = \partial_\mu\left[ \frac{\delta{\cal L}}{\delta \partial_\mu\phi} -\partial_\nu \frac{\delta{\cal L}}{\delta \partial_{\nu}\partial_{\mu}\phi}\right]=0$. The current has a single nonvanishing component, $J^\mu= J_0\delta^\mu_z$ given by $J_0= a^3 j \psi ={\rm const.}$ with
\begin{equation}\label{e.EMphi0}
 j = {\cal K}' - 6 H_{\rm E}^2(G'_4 + 2 X_{\rm E}^2 G''_4) =\frac{J_0}{a^3\sqrt{X_{\rm E}}}\,.
\end{equation}
Then, the variation of the action with respect to the metric leads to two equations, respectively for  $N_{\rm E}$ and $h$,
\begin{eqnarray}
&& {\cal K}-2 X_{\rm E} {\cal K}' - 6 H_{\rm E}^2\left[ G_4 -4X_{\rm E} G'_4 - 4X_{\rm E}^2 G_4''\right]=0\,,\quad\label{e.evoHN1} \\
&&{\cal K} -2 \left(3H_{\rm E}^2+2\frac{\dot{H}_{\rm E}}{N_{\rm E}}\right)G_4 \nonumber \\
&& \qquad+ 4\left[\left(3H_{\rm E}^2+2\frac{\dot{H}_{\rm E}}{N_{\rm E}}\right)X_{\rm E}+ H_{\rm E}\frac{\dot{X}_{\rm E}}{N_{\rm E}}\right] G_4'\quad \nonumber \\
&& \qquad+ 8H_{\rm E}\frac{\dot{X}_{\rm E}}{N_{\rm E}} X_{\rm E} G_4''  =0\, .\label{e.evoHS1}
\end{eqnarray}
Indeed only two of the three equations~(\ref{e.EMphi0}-\ref{e.evoHS1}) are independent thanks to the Bianchi identity. Hence, for the sake of simplicity, we only use the first two~(\ref{e.EMphi0} and \ref{e.evoHN1}) in our analysis. Defining
\begin{equation}\label{e.defcalG}
{\cal G}\equiv G_4-2 X_{\rm E} G'_4 = -2X_{\rm E}^{3/2}\left(\frac{G_4}{\sqrt{X_{\rm E}}} \right)' \,,
\end{equation}
they simplify to
\begin{equation} \label{e.systFin}
{\cal K}'+6H_{\rm E}^2{\cal G}'=\frac{J_0}{a^3\sqrt{X_{\rm E}}}\,,
\quad
{\cal K}-6H_{\rm E}^2{\cal G}=2\frac{ J_0 \sqrt{X_{\rm E}}}{a^3}\,,
\end{equation}
a system which describes the whole theory and is independent of the choice of $N_{\rm E}$.

\subsection{Necessary condition for the existence of solutions}\label{subsec2.2}

Equation~(\ref{e.EMphi0}), or the first of (\ref{e.systFin}), has an important and generic implication on the field profile.

Suppose that $\psi(z=z_0)=0$ and that $j(z=z_0)$ is finite and nonzero. Then $J_0=0$ and thus $j(z)\psi(z) = 0$ for all $z$. Also, by continuity $j(z)$ is finite and nonzero in an open neighborhood of $z=z_0$. These two facts imply that $\psi(z)=0$ in the open neighborhood, meaning in particular that $\psi$ and its all-order derivatives vanish at $z=z_0$. Since $\phi(z)$ satisfies a second-order ordinary differential equation and $\psi(z)$ is its first derivative, it is concluded that $\psi(z)=0$ for all $z$. Therefore, $\psi$ cannot cross zero unless $j$ also vanishes there. 

To evade this no-go result, we can for example consider clock-field configurations with ``$Z_2$ antisymmetry,'' {\em i.e.},
\begin{equation}
 \phi(-z)-\phi_0 = -[\phi(z)-\phi_0]\,
\end{equation}
with $\phi_0$ an arbitrary constant, and metric configurations with ``$Z_2$ symmetry,''
\begin{equation}
 a(-z) = a(z)\,, \quad N_{\rm E}(-z) = N_{\rm E}(z)\,.
\end{equation}
In particular, this implies
\begin{equation}
 \psi(-z) = \psi(z)\,, \quad H_{\rm E}(-z) = -H_{\rm E}(z)\,,
\end{equation}
and
\begin{equation}
 \dot{\psi}(z=0) = 0\,, \quad H_{\rm E}(z=0) = 0\,.
\end{equation}
In such a case, the existence of a nontrivial solution with $j\to 0$ ($z\to\pm\infty$) is not forbidden by the no-go and could be found by tuning $\psi(z=0)$. If $H_{\rm E}(z)>0$ for $z>0$ then this is basically what we could call a Euclidean bouncing universe. The shift charge density $j \psi$ will automatically approach zero as $z\to\pm\infty$ if $\psi(z=0)$ is in the basin of the attractor. 

For the emergence mechanism to work, one needs that
\begin{equation}
g_{00} = (1-\psi^2/M^4)\,N_{\rm E}^2
\end{equation}
changes sign on the hypersurface $\Sigma_0$ defined by $g_{00}(\psi_c)=0$, {\em i.e.},
\begin{equation}\label{e.defpsic}
\psi(z_c)=\psi_c=M^2\equiv\sqrt{X_c} \,. 
\end{equation}
This happens if
\begin{equation} \label{e.psi02<M4<psiinfty2}
\psi^2(z=0) < M^4 < \psi^2(z=\pm\infty)\,
\end{equation}
or if
\begin{equation}
\psi^2(z=\pm\infty) < M^4 < \psi^2(z=0)\,.
\end{equation}
The previous argument teaches us that one has to set boundary conditions such that $\psi$ tends toward either zero or a constant that makes $j$ vanish as $z\to\pm\infty$ and that they never cross $\psi=0$. In order to have two external zones with Lorentz signature and a de Sitter geometry at $z=\pm\infty$, one needs to ensure that Eq.~(\ref{e.psi02<M4<psiinfty2}) holds and that $H_{\rm E}$ approaches a positive constant as $z\to\pm\infty$ . 

\subsection{Cosmological singularity}\label{subsec2.3}

We now analyze the cosmological solution that emerges in the region ${\cal M}_0$ where $g_{00}$ becomes negative and in which the metric is of the Friedmann-Lema\^{i}tre form
\begin{equation}
\mathrm{d} s^2 = -\mathrm{d} t^2 + R^2(t) \delta_{ij}{\rm d} x^i {\rm d} x^j\,.
\end{equation} 
Setting $N_{\rm E}=1$, the emergent cosmic time follows from
\begin{equation}\label{e.dtdz}
\mathrm{d} t^2 = -g_{00} \mathrm{d} z^2=\left(\frac{\psi^2}{M^4} -1\right)\mathrm{d} z^2
\end{equation}
and the scale factor is given by
\begin{equation}\label{e.rt}
 R(t) = a[z(t)]\,,
\end{equation}
hence providing the cosmological dynamics, {\em i.e.}, $R(t)$ and $H(t)=\partial_t\ln R(t)$ in the emergent region ${\cal M}_0$. It is important to note that Eq. \eqref{e.dtdz} implies that $H$ and $H_{\rm E}$ differ by a singular factor,
\begin{equation}\label{e.H2HE}
H = \left(\frac{\psi^2}{M^4} -1\right)^{-1/2}H_{\rm E}\,.
\end{equation}
It follows that while the scale factor remains on $\Sigma_0: \lbrace\psi^2=M^4\rbrace$, $H$ will be singular on $\Sigma_0$ if $H_{\rm E}$ remains nonvanishing. 

The singular behavior in $H$ should not be surprising, as the effective metric becomes degenerate on the hypersurface $\Sigma_0$. It has been pointed out in the literature that while a degenerate metric may correspond to coordinate singularities, a degenerate metric on a surface of codimension one may signal the presence of a distributional or thin-shell~\cite{Feng:2023vqp} singularity (cf. also Refs.~\cite{Ellis:1991sp,Ellis:1991st,Hayward:1992zp,Dray:1996cw,Dray:1996dc}) and other types of curvature singularities (as in the example presented in Ref.~\cite{Feng:2023klt}). In our scenario the hypersurface on which the effective metric becomes degenerate is completely regular from the viewpoint of the Euclidean theory and corresponds to the cosmological singularity from the viewpoint of the emergent spacetime. 
 
\subsection{General behavior}\label{subsec2.4}

To see general behavior of the system, it is convenient to rewrite the system~\eqref{e.systFin} in the following manner:
\begin{equation} \label{e.eomaH}
    H_{\rm E}^2=\frac{\mathcal{K}-2 X_{\rm E} \mathcal{K}'}{6\mathcal{G}+12X_{\rm E}\mathcal{G}'}\,,
    \quad 
    a^3 = \frac{J_0 (\mathcal{G}+2X_{\rm E}\mathcal{G}')}{\sqrt{X_{\rm E}}(\mathcal{G}\mathcal{K})'}\,.
\end{equation}
Up to sign changes, $H_{\rm E}$ and $a$ are thus expressed algebraically as functions of $X_{\rm E}$. One can see that these equations implicitly specify the dependence of the system on $z$ through the relation $(\ln a^3)^.=3H_{\rm E}$. One can obtain a differential equation for $\psi$ by differentiating $a^3(X_{\rm E})$ with respect to $\psi$, employing the chain rule $\dot F= \dot X_{\rm E} F' = 2\psi\dot\psi F'$ to obtain
\begin{equation} \label{e.dpsidH}
    \frac{\dot{\psi}}{N_{\rm E}} =\frac{3 a^3 H_{\rm E}}{2\psi(a^3)'}\, ,\qquad \frac{\dot H_{\rm E}}{N_{\rm E}} = \frac{3 a^3}{2}\frac{(H_{\rm E}^2)'}{(a^3)'}\,.
\end{equation}
The expression for the derivative $\dot H_{\rm E}/N_{\rm E}$ is obtained from the chain rule and the expression ${\dot{\psi}}/{N_{\rm E}}$. Observe that the rhs of each expression in Eq.~\eqref{e.dpsidH} may be expressed in terms of $\psi=\sqrt{X_{\rm E}}$ using the expressions in Eq.~\eqref{e.eomaH}. This allows us to extract the general behavior of the system from the functional forms of $\mathcal{K}(X_{\rm E})$ and $\mathcal{G}(X_{\rm E})$ without solving the differential equations. Of course, if the rhs of Eq.~\eqref{e.eomaH} is expressed explicitly in terms of $\psi$, one may also read it as a first-order differential equation for $\psi(z)$; given the solution $\psi(z)$, the algebraic expressions~\eqref{e.eomaH} supply the $z$ dependence (up to sign changes) for $H_{\rm E}$ and $a$.

\subsection{Simple example}\label{subsec2.5}

\begin{figure}[htb]
	\centering
	\includegraphics[width=0.23\textwidth]{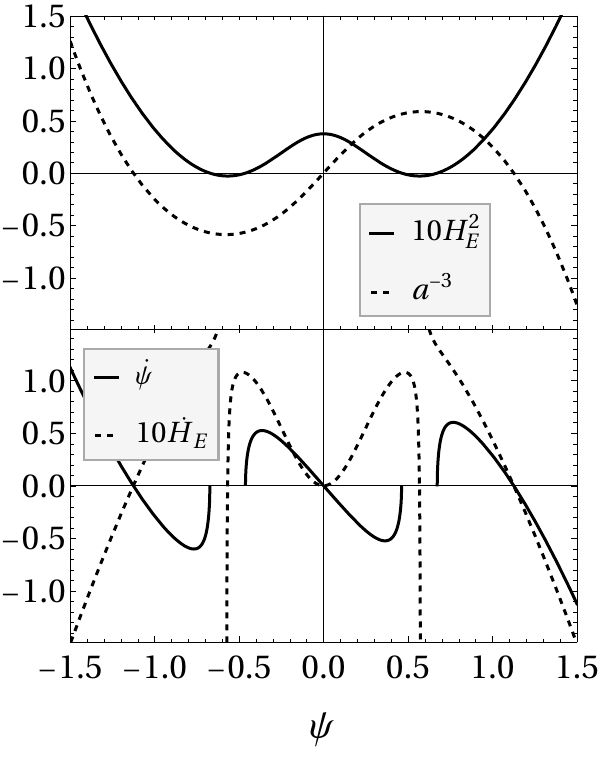}\hskip.1cm\includegraphics[width=0.23\textwidth]{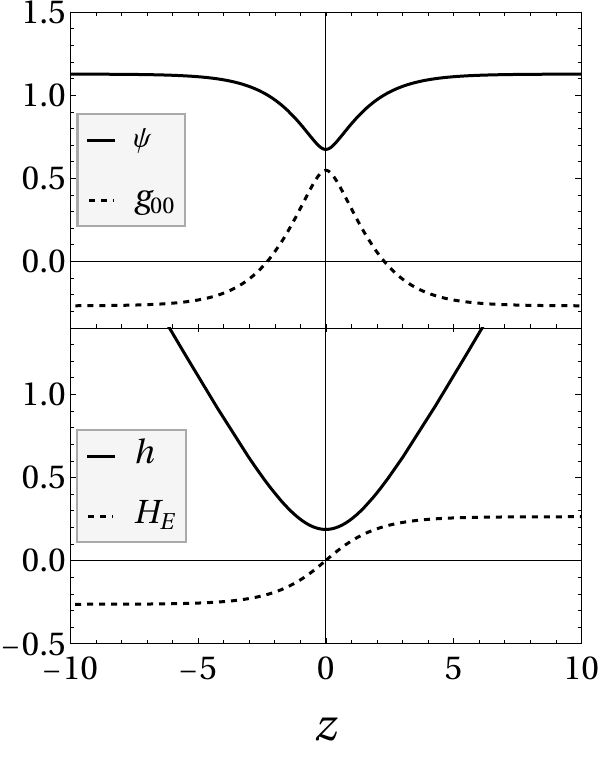}
	\caption{Integration of the field equations assuming $G_4$ and ${\cal K}$ of the form~(\ref{e.choice}) with parameters $(g_0,g_1,k_0,k_2,X_0)=(1, -1, 1, -1.55, 1)$ and the integration constant $J_0=1$. The upper line shows the intermediary functions $(10H_{\rm E}^2,a^3)$ [left] as a function of $\psi$ obtained from Eq.~(\ref{e.eomaH}) and $(10 \dot{H}_{\rm E},\dot\psi)$ [right] as a function of $\psi$ obtained thanks to Eqs.~\eqref{e.dpsidH}. The missing segments in the plot of $\dot\psi$ correspond to regions where $H_{\rm E}^2<0$. The lower plot gives the full solution for the clock field $\psi$ and the geometry $H_{\rm E}$ as functions of $z$.}
	\label{fig:LinG4}
\end{figure}

To give a concrete example, we consider a specific class of ${\cal K}$ and $G_4$. In the early work~\cite{Mukohyama:2013gra} a part of the Lagrangian of the renormalizable scalar-tensor theory contains ${\cal K}$ that is quadratic in $X_{\rm E}$ and $G_4$ that is linear in $X_{\rm E}$.  This led us to assume that
\begin{eqnarray}\label{e.choice}
G_4 = g_0 + g_1 X_{\rm E}, \quad \mathcal{K} = k_0 + \tfrac{1}{2}k_2 (X_0-X_{\rm E})^2.
\end{eqnarray}
The result of the procedure described in Sec.~\ref{subsec2.4} is summarized in Fig. \ref{fig:LinG4} that shows $H_{\rm E}^2$, $a$, $\dot\psi$ and $\dot H_{\rm E}$ as functions of $\psi$ and then $\psi$ and $H_E$ as functions of $z$, where we have set $N_{\rm E}=1$. The profile of the clock field allows, depending on the mass scale $M$, for the apparition of two Lorentzian universes respectively for $z>z_c$ and $z<-z_c$ with $z_c$ the positive root of Eq.~(\ref{e.defpsic}). Then at large $z$, $\psi$ tends to a constant so that $t\sim z$ and since $H_E$ also tends to a constant, $H$ tends to a constant at large $t$. Hence, this allows us to conclude that two {\em mirror} emergent universes with a proper primordial almost de Sitter phase and ``emergent time flowing in different directions'' can be obtained with the simple choice of the free functions.

\section{Reconstruction}\label{subsec3}

\subsection{Formalism}\label{subsec3.1}

The system~(\ref{e.systFin}) can be seen as a set of equations to determine $\lbrace {\cal K},{\cal G} \rbrace$ from a given solution, {\em i.e.}, $\lbrace a(z),\psi(z)\rbrace$ from  which one can, at least formally extract $X_{\rm E}(z)$, $H_{\rm E}(z)$ and $h(z)$ as well as $a({\rm X_E})$ and $H_{\rm E}(X_{\rm E})$, provided that $N_{\rm E}$ is fixed ({\em i.e.}, $N_{\rm E}=1$ in the following).

First, the two equations~(\ref{e.systFin}) can be combined to get $(H_{\rm E}\mathcal{G})'=J_0 \sqrt{X_E} a' /2 H_{\rm E} a^4$ from which one obtains the formal integral solution
\begin{equation} \label{eq:EoMsCombinedsolformal}
    \mathcal{G}(X_E) 
    = 
    \frac{1}{H_{\rm E}(X_{\rm E})} 
    \left[C_0 + \frac{J_0}{2} \int_{0}^{X_{\rm E}} \frac{\sqrt{\tilde{X}_{\rm E}}}{H_{\rm E}(\tilde{X}_{\rm E})}\frac{a'(\tilde{X}_{\rm E})}{a^4(\tilde{X}_{\rm E})} \mathrm{d} \tilde{X}_{\rm E}\right]\,,
\end{equation}
from which one easily deduces $\mathcal{K}$ from the second equation of Eq.~(\ref{e.systFin}) as
\begin{equation} \label{eq:Krecon}
    \mathcal{K}(X_{\rm E})=6H_E^2 \mathcal{G}(X_{\rm E})+2 J_0 \frac{\sqrt{X_{\rm E}}}{a^3}\,
\end{equation}
and then $G_4$ by a direct integration of Eq.~(\ref{e.defcalG}). 

While this formally demonstrates the possibility of the reconstruction, this procedure is not practical unless $X_{\rm E}(z)$ can be inverted analytically since $a$ and $H_{\rm E}$ are not directly known as a function of $X_{\rm E}$ but as a function of $z$. This is easily circumvented by noting that whatever function  $F$, $\dot F = F' \times \dot X_{\rm E}$. Hence $(H_{\rm E}\mathcal{G})^{\cdot}/N_{\rm E}=J_0 \sqrt{X_E} /2 a^3$ so that Eq.~(\ref{eq:EoMsCombinedsolformal}) becomes
\begin{equation} \label{e.calGofz}
    \mathcal{G}(z) 
    = 
    \frac{1}{H_E(z)} 
    \left[C_0 +\frac{J_0}{2}  \int_{0}^{z}  \frac{\psi(\tilde{z})}{a^3(\tilde{z})} N_{\rm E}(\tilde{z}) \mathrm{d} \tilde{z}\right]\,,
\end{equation}
which, once combined with $X_{\rm E}(z)$, trivially provides  $\mathcal{G}(X_{\rm E})$ in a parametric form. Then Eq.~(\ref{eq:Krecon}) gives
\begin{equation} \label{e.calKofz}
    \mathcal{K}(z)=6H_E^2(z) \mathcal{G}(z)+\frac{2 J_0 \sqrt{X_E}(z)}{a^3(z)}\,,
\end{equation}
and then $G_4(z)$ is obtained as
\begin{equation} \label{e.G4ofz}
    G_4(z)=\left[ C_1-  \int_0^{z} \frac{{\cal G}(\tilde{z})\dot\psi(\tilde{z})}{\psi^2(\tilde{z})} \mathrm{d} \tilde{z} \right] \sqrt{X_E}(z)\,.
\end{equation}
Note that whatever the value of the integration constant $C_1$, the expression~(\ref{e.G4ofz}) of $G_4$ leads to the same expression for ${\cal G}$ and hence to the same solution $\lbrace \psi(z), h(z)\rbrace$ of the system~(\ref{e.systFin}). This reflects the fact that one can always add a term proportional to $\sqrt{X_E}$ to $G_4(X_E)$ without changing $\mathcal{G}(X_{\rm E})$. Hence, we shall set $C_1=0$ in the following numerical examples.

Let us also stress that the case $J_0=0$, in which the formal integrals simplify considerably, must be handled with care. In this case one reconstructs $\mathcal{G}(X_{\rm E})=C_0/\tilde{H}_{\rm E}(X_{\rm E})$ and $\mathcal{K}(X_{\rm E})=6C_0\tilde{H}_{\rm E}(X_{\rm E})$, where we have renamed the input $H_{\rm E}(X_{\rm E})$ as $\tilde{H}_{\rm E}(X_{\rm E})$. However, once the Lagrangian is reconstructed in this way, the set of equations (\ref{e.systFin}) for $\lbrace h(z)$, $\phi(z)\rbrace$ generically becomes ill behaved: ({\em i}) if $\tilde{H}_{\rm E}$ is not proportional to $\sqrt{X_{\rm E}}$ and if the algebraic equation $2\tilde{H}'_{\rm E}(X_{\rm E})X_{\rm E}=\tilde{H}_{\rm E}(X_{\rm E})$ does not admit a non-negative real solution, then there is no solution with $J_0\ne 0$; ({\em ii}) if $\tilde{H}_{\rm E}$ is not proportional to $\sqrt{X_{\rm E}}$ and if the algebraic equation admits a non-negative real solution $X_{\rm E}=X_{{\rm E}*}=const.$ then the only solution with $J_0\ne 0$ is those with $\psi(z)=\pm\sqrt{X_{{\rm E}*}}$ and thus generically exhibits discontinuity at $J_0=0$ where $\psi(z)$ is undetermined by the equations of motion; ({\em iii}) if $\tilde{H}_{\rm E}$ is proportional to $\sqrt{X_{\rm E}}$ then $\psi(z)$ is undetermined by the equations of motion. 

\subsection{Worked out example and implications}\label{subsec3.2}

\begin{figure}[htb]
 	\centering
     \includegraphics[width=0.22\textwidth]{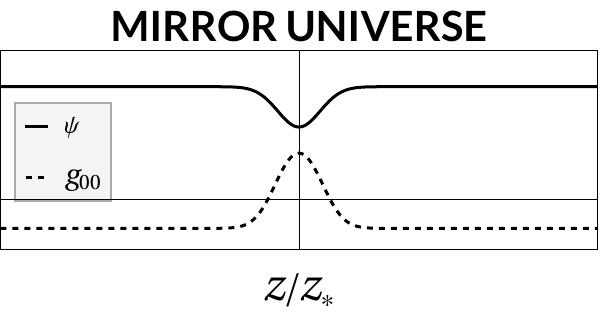}\hskip.1cm\includegraphics[width=0.22\textwidth]{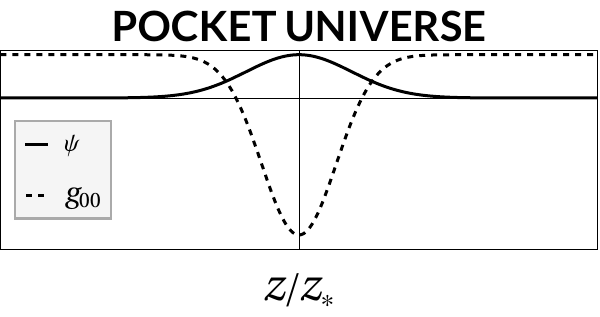}
      \includegraphics[width=0.45\textwidth]{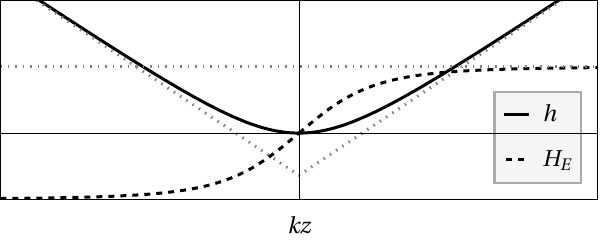} 
 	\caption{The considered solutions to be reconstructed require to specify the profile of the geometry $h$ (solid line) and $H_{\rm E}$ (dashed line) [bottom] as given by Eq.~(\ref{e.sol2}) and of the clock field $\psi$ [top] as given by Eq.~(\ref{e.sol1}) for a model of {\em mirror universes} [left] and {\em pocket universe} [right]. Asymptotically at large $z$, $H_{\rm E}\rightarrow$~const. and $h\sim H_{\rm E} z$ while $\psi\rightarrow const.$, allowing for the emergence of a de Sitter phase at late time in the case of the mirror universe.}
 	\label{fig:solution}
 \end{figure}

\begin{figure}[htb]
 	\centering
\includegraphics[width=0.45\textwidth]{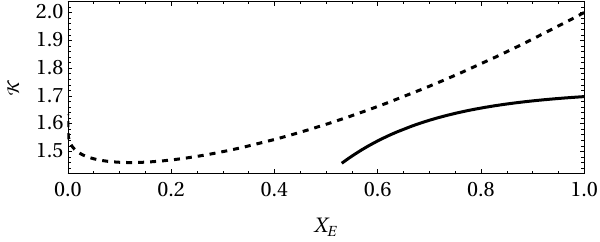}
\vskip0.25cm
\includegraphics[width=0.45\textwidth]{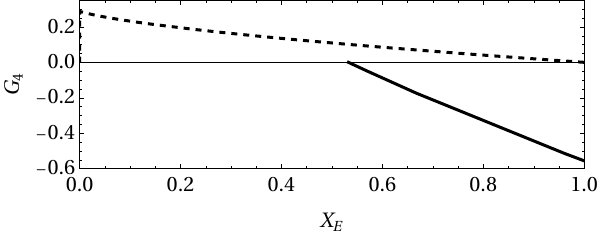}
 	\caption{Reconstruction of the free functions ${\cal K}(X_{\rm E})$ [top] and $G_4(X_{\rm E})$ [bottom] for a {\em mirror universe} [solid lines] and {\em pocket universe} [dashed lines] assuming $J_0=1$ and $C_0=0$. The mirror universe curve is not defined below the value of $X_E$ corresponding to the minimum value of $\psi$ occurring at $z=0$ in Fig. \ref{fig:solution}, and from the form of Eq. \eqref{e.sol1}, one can see that the values of $\psi$ and consequently $X_E$ are bounded.}
 	\label{fig:model}
 \end{figure}

\begin{figure}[htb]
 	\centering
\includegraphics[width=0.4\textwidth]{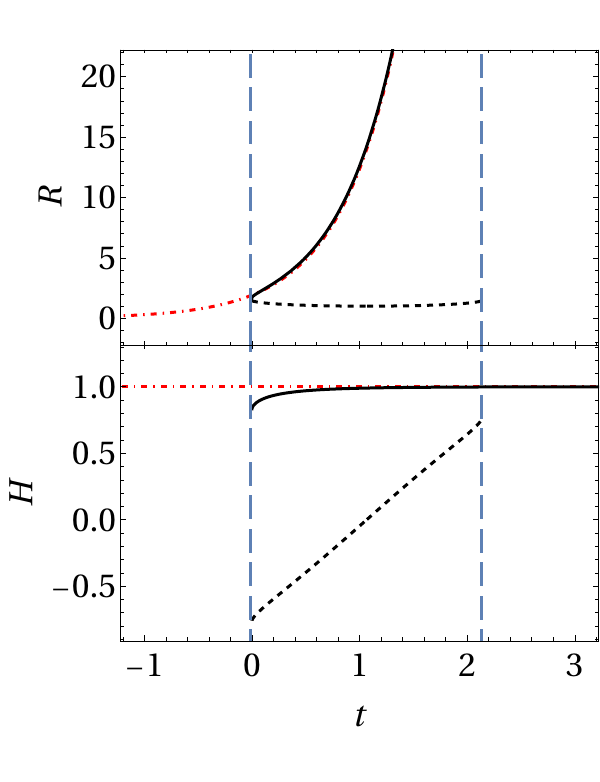}
 	\caption{The emergent cosmologies for {\em mirror universes} [solid lines] and {\em pocket universe} [dashed lines]. The scale factor $R$ and the Hubble rate $H$  as a function of the cosmic time $t$---assuming $t=0$ on $\Sigma_0$---are respectively plotted on the top and bottom frames for the same parameters as in Fig.~\ref{fig:model}. The red dot-dashed curve represents an exact de Sitter spacetime and the vertical dashed lines mark the limit of the Lorentzian regions ${\cal M}_0$. Contrary to the {\em mirror universes}, a {\em pocket universe} has a boundary in the future and hence a finite maximum age. }
 	\label{fig:cosmo}
 \end{figure}

The previous section has exhibited a model depicted in  Fig.~\ref{fig:LinG4} with two emergent {\em mirror universes} connected by a Euclidean region. The reconstruction approach allows to illustrate the variety of solutions that can in principle exist in the theory. To that purpose we assume that the spatial geometry~(\ref{e.spacegeo})  behaves as
\begin{equation}\label{e.sol2}
h(z)= 2 k z \arctan(k z) /\pi\,,
\end{equation}
which has the property that $h\sim k z$ at large $\vert z\vert$ so that $H_{\rm E}(z)\rightarrow \pm k$. Then, two kinds of configurations can be constructed depending on the clock-field profile, either {\em mirror universes}, as already obtained, or {\em pocket universes}  embedded in a Euclidean sea, a simple choice being
\begin{equation}\label{e.sol1}
\psi(z)=\left\lbrace
\begin{array}{ll}
\psi_0-\Delta\psi\exp\left(-z^2/z_*^2\right) & \hbox{[Mirror]} \\
\exp\left(-z^2/z_*^2\right) & \hbox{[Pocket]}
\end{array}\,,
\right.
\end{equation}
with $\Delta\psi<\psi_0$ and $\psi_0>0$ to ensure that $\psi$ does not cross 0, as imposed by the analysis of Sec.~\ref{subsec2.2}. Figure~\ref{fig:solution} illustrates these two classes of models and Fig.~\ref{fig:model} depicts the reconstructed free functions $\lbrace {\cal K}(X_{\rm E}), G_4(X_{\rm E})\rbrace$ that depend on the two integration constants $(J_0,C_0)$ assuming, as discussed above, that $C_1=0$. To finish, we give the emergent cosmologies, that is $R(t)$ and $H(t)$ in Fig.~\ref{fig:cosmo}.

This demonstrates that there exist solutions of the two families. Concerning the {\em mirror universes}, the expansion starts from a finite value of the scale factor at $t=0$ on $\Sigma_0$ and expands indefinitely while converging to a de Sitter behavior. Interestingly, due to the variation of $\psi$ close to $\Sigma_0$, the universe has a primordial phase prior to its de Sitter phase. This can eventually have cosmological signatures the study of which is beyond the scope of this paper.  Concerning the {\em pocket universes}, it has a finite lifetime since there is a boundary $\Sigma_0$ in the future. Since $\psi$ has no time to settle to constant value, the scale factor never behaves as de Sitter but enjoys a mild bounce.

\section{Discussion}\label{section3}

This paper gives a proof of concept that within the framework for the emergence of the Lorentz signature originally proposed by Refs.~\cite{Mukohyama:2013ew,Kehayias:2014uta}, one can construct pocket universes with emergent dynamical expansion of Friedmann-Lema\^{\i}tre kind. In this approach, the standard big bang singularity is replaced by a three dimensional hypersurface $\Sigma_0$ on which the signature of the metric flips. Hence, dynamics freezes outside the Lorentzian pockets that eventually emerge. It should be emphasized that the Euclidean metric $g^{\rm E}_{\mu\nu}$ and the scalar clock field $\phi$ remain regular at $\Sigma_0$, so it is not a singularity in these variables.

Our analysis relies on the action~(\ref{e.S1}) initially introduced in Ref.~\cite{Mukohyama:2013ew}. Our construction also assumed a planar symmetry in order to recover a spatially homogeneous and isotropic spacetime in the Lorentzian regions, as depicted in Fig.~\ref{fig:1}. But this can only be thought as a local approximation since  this symmetry can {\em a priori} not be extended over the whole four-dimensional space. Hence, our Universe may also have a spatial boundary on scale larger than our observable Universe.

 \begin{figure}[htb]
 	\centering
 	\includegraphics[width=0.5\textwidth]{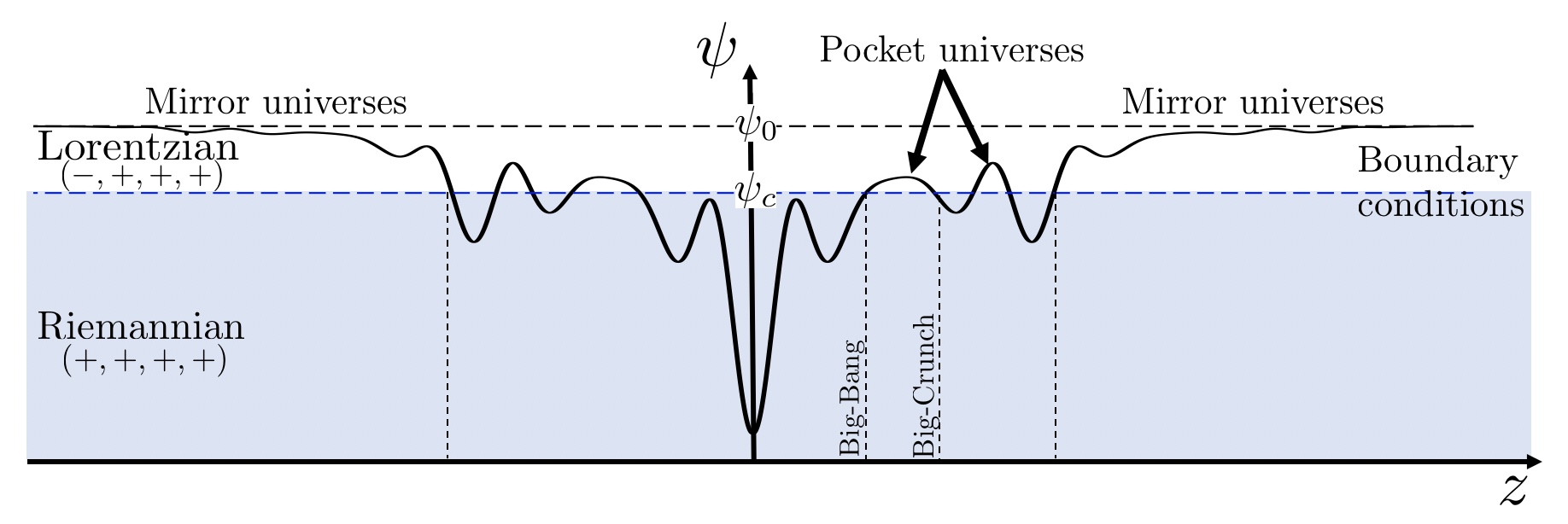} 
 	\caption{Once boundary conditions are set, the profile of the clock field cannot cross the line $\psi=0$ as demonstrated in Sec.~\ref{subsec2.2}. $\psi_c=M^2$ defines the change-of-signature hypersurface $\Sigma_0$ and delimitates the Lorentzian and Euclidean regions; see Eq.~(\ref{e.defpsic}). $\psi_0$ is defined by the condition $j(\psi_0)=0$; see Eq.~(\ref{e.EMphi0}). From a generic point of view, we may expect to have a nonmonotonic profile so that the four dimensional- Riemannian space may contain many Lorentzian ``pockets'' that may or not be bounded in the $z$, {\em i.e.} emergent time, direction.}
 	\label{fig:2b}
\end{figure} 

\begin{figure}[htb]
 	\centering
 	\includegraphics[width=0.5\textwidth]{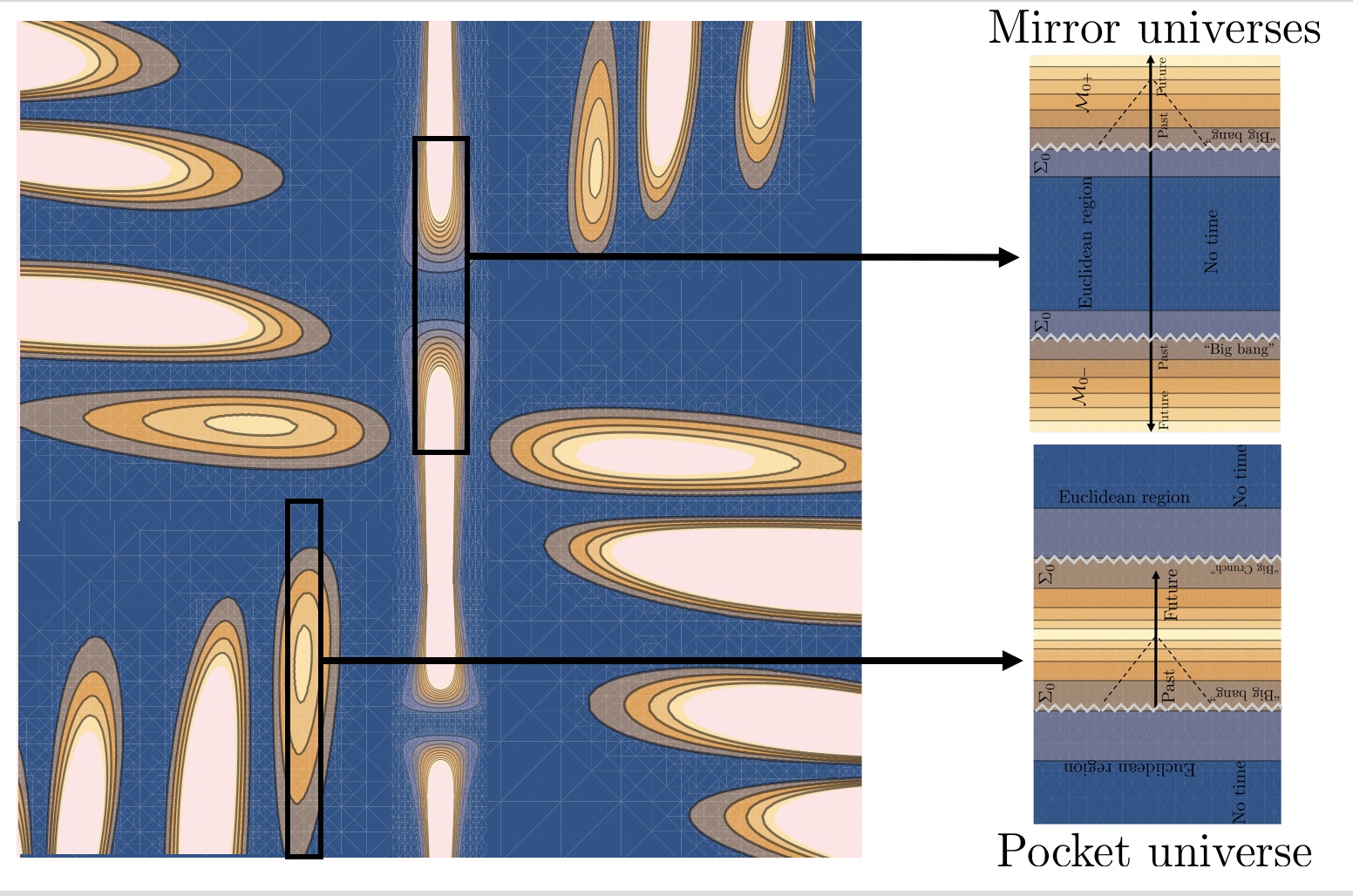} 
 	\caption{A big bang that never happened.... with many emergent universes. The dashed lines denote the light cone of an observer in ${\cal M}_{0+}$ for whom the whole structure remains unknown and inaccessible. Not only are those regions out of its domain of influence but also in the blue Riemannian sea in which time does not flow. According to the field distribution and the Euclidean metric, we can describe many types of universes: {\em mirror universes} and {\em pocket universes} of different size and lifetime.}
 	\label{fig:3}
 \end{figure} 

First, we have derived in~Sec.~\ref{subsec2.2} a necessary condition for a Lorentzian pocket to emerge, in the form of a no-go theorem. We have then studied in~Sec.~\ref{subsec2.4} the cases that evade the no-go and showed an explicit example in which a Euclidean island is embedded between universes with the Lorentzian signature. In this case the emergent geometry in each of the two asymptotic regions enjoys an  almost de Sitter expansion while the initial cosmological singularity of each universe is replaced by the hypersurface $\Sigma_0$ on which the effective metric becomes degenerate and beyond which it has the signature $(+,+,+,+)$. Then~Sec.~\ref{subsec3} has demonstrated that the reconstruction procedure can be explicitly solved, {\em i.e.}, one can determine the two free functions $G_4$ and ${\cal K}$ from the choice of the clock field and spatial geometry profiles, that is from $\psi(z)$ and $a(z)$. This approach allowed us to show that our model contains a rich phenomenology containing at least to classes of emergent spacetime: {\em mirror universes} already discussed in~Sec.~\ref{subsec2.4} and {\em pocket universes}; see Fig.~\ref{fig:model}. In the latter case, the Friedmann-Lema\^{\i}tre spacetime exhibits a bounce but the dynamics is bounded by $\Sigma_0$ both in the future and the past; see Fig.~\ref{fig:cosmo}. 

While beyond the scope of this first analysis, it is clear that in principle one could mimic the usual cosmological big bang expansion history $a_{\rm BB}(t)$ after a choice of clock-field profile $\psi(z)$. Indeed, imagine that $\psi$ becomes linear outside of a neighborhood of $\Sigma_0$ -- $\psi(\vert z\vert> z_0) \sim M\vert z\vert$ -- then Eq.~(\ref{e.dtdz}) is integrable as
$$
t(z)= \frac{z}{2}\sqrt{\frac{z^2}{M^2}-1} -M \ln\left[\frac{z}{M} +\sqrt{\frac{z^2}{M^2}-1}\right]
$$
from which it follows that $h(z)$ is given by $\ln a_{\rm BB}[t(z)]$ and the reconstruction procedure ensures that in this {\em ad hoc} setting, a model reproducing the whole expansion history of our observable Universe can be constructed. Indeed, such an argument of existence is far from being sufficient since we shall consider free functions that are fixed and study {\em e.g.}, whether this requires a fine-tuning and the behavior of solutions with other symmetries, etc.

Even if this new cosmological construction may seem very speculative, it offers a new vision of the primordial Universe and of the Universe on scales larger than those of our observable Universe, which gives a concrete reconstruction of the early insight by Sakharov~\cite{Sakharov:1984csx}. Besides, it has specific properties and signatures. {\em First}, the big bang is no more a singularity in the fundamental (Euclidean) variables and $R(t=0)$ remains finite as seen in Fig.~\ref{fig:solution}. {\em Second}, the expansion in the primordial phase may not be directly de Sitter so that inflation would be preceded by an expansion phase that could let a specific signature on the primordial fluctuations. Indeed, this remains to be investigated with care. {\em Third}, the physical extension of the three-dimensional spatial ({\em i.e.}, constant $t$ or $z$) hypersurfaces may be bounded so that one may have a finite universe without invoking a spatial topology. Indeed it has to be larger than the observable Universe but may still leave detectable signatures, {\em e.g.} on CMB temperature and polarization distributions~\cite{Fabre:2013wia}. {\em To finish}, there shall be gravity beyond general relativity and the amplitude of the deviations from general relativity can be predicted once a full model is constructed and its parameters determined by the cosmology, hence making the construction, at least in principle, falsifiable. The propagation of test fields in spacetime with Euclidean pockets has already been studied in analogue systems \cite{Weinfurtner:2007br,Donley:2001wvu,Weinfurtner:2007dq,Nissinen:2017}, which can be a setting for exploring primordial fluctuations.

It is also worth pointing out that the change in the metric signature of the effective metric $g_{\mu\nu}$ fundamentally modifies the nature of the matter field equations since they shift from elliptic to hyperbolic---as was actually the early motivation of Refs.~\cite{Mukohyama:2013ew,Kehayias:2014uta,Mukoyama:2013gqu,Mukohyama:2013gra,Uzan-CUP}. While a Lorentzian effective metric is required at long distance scales, on short distance scales in the vicinity of $\Sigma_0$, one will need to introduce higher derivative terms in the matter action coupled to $g^{\rm E}_{\mu\nu}$ in order to avoid the potential pathologies that arise from coupling to a signature changing metric.  In a similar manner, a more complete treatment of the gravitational sector will require a higher derivative model as well, following that introduced in Ref.~\cite{Mukohyama:2013gra} and Eq.~(2.5) therein, {\em i.e.},
\begin{eqnarray}\label{e.S2}
S &=& \int  \sqrt{g_{\rm E}} \left\lbrace G_4(X_{\rm E})R_{\rm E}  - \frac{\omega}{3\lambda} R^2_{\rm E} +\frac{1}{2\lambda}C^2_{\rm E} \right. \\
&& \left. +{\cal K}(X_{\rm E} ) + \alpha (\nabla^2\phi)^2 +\beta (\nabla_\mu^{\rm E}\nabla_\nu^{\rm E}\phi)^2\right\rbrace \mathrm{d}^4x \nonumber 
\end{eqnarray}
with ${\cal K}(X_{\rm E} ) = (2Z\Lambda_{\rm E} - 2X_*X_{\rm E} +X^2_{\rm E} )$ and $G_4(X_{\rm E} )= (\gamma X_{\rm E} - Z)$ so that it has a modified coupling to the scalar field since one allows for $\beta\not=\alpha$ besides other higher derivative terms. Ultimately the stability analysis of inhomogeneous perturbations should also be performed in these higher derivative theories with proper boundary conditions, as in Ref.~\cite{Feng:2025xsi}. In the present work, we have only shown a proof of concept for the emergent big bang scenario at the level of the plane-symmetric background. 

To conclude, we can grasp the big picture on the Universe at large that emerges when  the description developed in this paper. It is illustrated in Figs.~\ref{fig:2b} and~\ref{fig:3}. Since the large-scale distribution of the clock field shall depend on boundary conditions and eventually on the geometry of the four-dimensional Euclidean space, one expects it to be nontrivial. In full generalities, many Lorentzian pockets can emerge with their own geometry and with a large span of sizes, {\em i.e.} of expansion time as seen from their inside. It also follows that the description of the primordial Universe may not require it to go through a quantum gravity phase even if it requires gravity beyond general relativity. This proposal opens a new, rich and exciting phenomenology that needs to be investigated further and that can eventually change our conception of the cosmos on the largest scales.

\begin{acknowledgments}
We are deeply grateful to Sante Carloni for helpful discussions during the early stages of this work.
J. C. F. is grateful for the hospitality of the Research Center for the Early Universe (RESCEU) and University of Tokyo, as well as the Yukawa Institute for Theoretical Physics (YITP), Kyoto University, during which part of this work was performed. J. C. F. is supported by the European Union and Czech Ministry of Education, Youth and Sports through the FORTE Project No. CZ.02.01.01/00/22\_008/0004632. The work of S. M. was supported in part by JSPS (Japan Society for the Promotion of Science) KAKENHI Grant No.\ JP24K07017 and World Premier International Research Center Initiative (WPI), MEXT, Japan. J.-P. U. is thankful to YITP for hospitality during his visit.
The Mathematica file used to perform the numerical integration is available 
in the GitHub repository~\cite{Feng:2026mat}.
\end{acknowledgments}

\bibliographystyle{bib-style}

\end{document}